\documentclass[prereview, 5p]{elsarticle}

\usepackage[numbers]{natbib}
\usepackage{amsmath}
\usepackage[hyphens]{url}
\usepackage{hyperref}
\begin{document}
\begin{frontmatter}
\title{Accelerated Time-Domain Simulation of Complex Photonic Structures with a Data-Aware Fourier Neural Operator}

\author[1]{Zaifan Wu}
\author[1]{Yue You}
\author[2]{Xian Zhou}
\author[1,3]{Fan Zhang\corref{cor1}\fnref{fn1}} 

\ead{fzhang@pku.edu.cn}

\cortext[cor1]{Corresponding author: Fang Zhang (fzhang@pku.edu.cn)}

\affiliation[1]{organization={State Key Laboratory of Photonics and Communications, 
                              School of Electronics, Peking University},
                city={Beijing},
                postcode={100871},
                country={China}}

\affiliation[2]{organization={School of Computer \& Communication Engineering, 
                              University of Science and Technology Beijing},
                city={Beijing},
                postcode={100083},
                country={China}}

\affiliation[3]{organization={Peng Cheng Laboratory},
                city={Shenzhen},
                postcode={518055},
                country={China}}

\begin{abstract}
Efficient and accurate time-domain simulation of electromagnetic fields in complex photonic devices is critical for designing broadband and ultrafast optical components, yet it is often limited by the high computational cost of conventional numerical methods like FDTD. While machine learning approaches show promise in accelerating these simulations, existing models still struggle to simultaneously capture the dynamic field evolution and generalize to complex geometries. In this paper, we introduce a Data-Aware Fourier Neural Operator (DA-FNO) as an innovative neural operator for solving electromagnetic simulations. Applied autoregressively, the model iteratively predicts the time-domain evolution of all field components and automatically terminates upon energy convergence. Our model not only generalizes to complex and randomized geometries but also shows good predictive consistency across the optical C-band (1530-1565nm) when evaluated on the test set. In a representative configuration, it achieves an 11$\times$ speedup over conventional methods while maintaining about 95\% accuracy across the C-band. This approach provides a new pathway for C-band photonic simulations, potentially facilitating the research, development, and inverse design of novel photonic devices.
\end{abstract}

\end{frontmatter}

\section{Introduction}
Solving Maxwell's equations remains a fundamental challenge in photonics. Current engineering practices predominantly rely on numerical methods such as the Finite Element Method (FEM) \cite{volakis_fem_1998,jin_fem_2015}, Finite Difference Frequency Domain (FDFD) \cite{zhao_fdfd_2002,hughes_fwdmode_2019}, and Finite Difference Time Domain (FDTD) \cite{taflove_fdtd_chapter_2005,taflove_fdtd_2013} algorithms. Although full-wave Maxwell solvers provide accuracy, their substantial computational costs restrict the scalability of applicable scenarios. Among these methods, the FDTD algorithm stands out for its explicit temporal evolution of electromagnetic fields, enabling efficient time-resolved simulations such as ultrafast optics and broadband excitation \cite{OQE1_2023,OQE2_2023,OQE3_2023}. However, its strict adherence to the Courant-Friedrichs-Lewy (CFL) stability condition imposes stringent spatiotemporal resolution constraints, leading to excessive computational costs for high-precision or large-scale simulations.

In recent years, machine learning (ML) approaches have been explored to overcome these computational bottlenecks. To capture the temporal evolution of fields, early efforts utilized recurrent architectures like LSTMs\cite{LSTM_Origin_1997} and RNNs combined with convolutions\cite{cnn_Origin_2012}. While successful in modeling time-resolved dynamics, these approaches were often limited in their ability to generalize beyond the simple, regular geometries on which they were trained\cite{yao_cnn_2018,noakoasteen_LSTM_CNN_2020,zhang_LSTM_2023,guo_RCNN_2023}. Other architectures like Transformers\cite{transformer_Origin,du_transformer_2023,noakoasteen_transformer_2024,lim_transformer_2024} and Graph Neural Networks (GNNs) have also been applied. For instance, GNNs have been used to perform a single-step FDTD update, but error accumulation prevented their extension to full iterative time stepping for long-term simulations\cite{GNN_APL_2023}.

\begin{figure*}[t!]
\centering
\includegraphics[width=469pt]{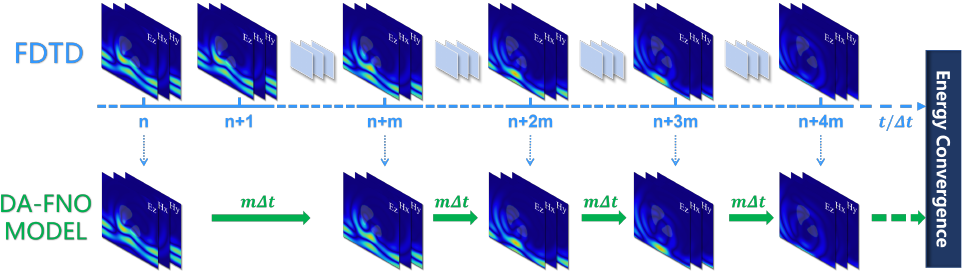}
\caption{Schematic diagram of the proposed model. Similar to the FDTD method, the DA-FNO model iteratively outputs all field components ($E_z$, $H_x$, $H_y$) over time until the energy converges. By eliminating CFL constraints, the DA-FNO allows for larger time steps ($m\Delta t$), thereby enabling faster simulations.}
\label{schematice_diagram}
\end{figure*}

Concurrently, two other paradigms have gained traction. Physics-Informed Neural Networks (PINNs)\cite{raissi_PINN_Origin_2019,PINN_Origin_JSC_2022} offer an unsupervised approach by embedding physical laws directly into the loss function, enabling PDE solving without labeled data. This has been applied to predict time-domain electromagnetic fields\cite{PINN_2D_TAP_2024,PINN_3D_TMTT_2025} and optimize nanophotonic structures like microlenses\cite{PINN_Maxwell_APL_2022,PINN_APL_2023} and metagratings\cite{chen_wavey-net_2022}. Separately, neural operators\cite{Neural_Operator_Origin_2020}, which learn mappings between function spaces, have shown great promise. Models like Fourier Neural Operators (FNOs)\cite{FNO_Origin_2021} and Deep Operator Networks (DeepONets)\cite{DeepONet_Origin} have been successfully applied to frequency-domain problems\cite{FNO_Hybrid_2023,jiang_DeepONet_2024}, such as modeling scattering from irregular structures\cite{FNO_ACS_2023}. A recent FNO-based model\cite{FNO_TAP_2025} effectively models time-domain electromagnetic fields over a fixed temporal horizon, producing predictions in a single forward pass.

Despite these advances, a critical gap remains. Most existing models are tailored to specific scenarios, and while some can handle irregular structures in the frequency domain, there is still no model that simultaneously achieves long-term time-domain simulation while generalizing to complex and irregular geometries.

In this work, we propose a Data-Aware Fourier Neural Operator (DA-FNO) to address this challenge, applied autoregressively to predict the evolution of 2D electromagnetic fields. The proposed model is capable of generalizing across unseen geometries, targeting Transverse Magnetic (TM)-polarized plane wave scattering in a simulation domain. Similar to the FDTD method, the model iteratively outputs all field components $(E_z,H_x,H_y)$ over time, allowing for dynamic tracking of the scattering process. Therefore, the DA-FNO model simultaneously supports the time-domain evolution of electromagnetic fields and generalization to complex and randomized geometries. In addition, it demonstrates strong wavelength generalization across the optical C-band. To further assess the model's scalability, we also conduct a preliminary investigation on 3D extruded geometries, verifying the model's capacity to learn volumetric features. The DA-FNO model monitors the instantaneous energy within the domain, progressively reducing energy as the outgoing scattered waves evolve, and automatically terminates the simulation once the energy falls below a convergence factor, $\delta$, of its historical maximum. Moreover, the elimination of CFL constraints enables the use of a coarse time-step ($m\Delta t$), creating a trade-off between computational speed and solution accuracy: increasing the step size (m) improves speed but decreases accuracy. In a representative configuration with $m=15$ and $\delta=10^{-2}$, the model attains an $11\times$ speedup while maintaining approximately 95\% accuracy. Fig.~\ref{schematice_diagram} illustrates the schematic diagram of the proposed model. The FDTD simulation is performed with a time step of $\Delta t$. The DA-FNO model outputs the three field components ($E_z,H_x,H_y$) of the TM-polarized wave at every $m\Delta t$ interval, enabling faster simulation compared to the FDTD. This process continues until the energy reaches convergence.

Full access to the implementation code and the first 200 training samples is provided.

\section{Model Construction}

\begin{figure*}[t]
\centering
\includegraphics[width=13cm]{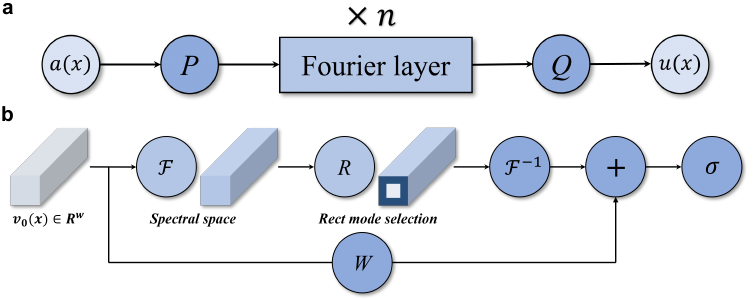}
\caption{(a) The full architecture of the vanilla FNO. $a(x)$ is lifted to a higher-dimensional space by network $P$, processed through $n$ Fourier layers and projected to the target function $u(x)$ by network $Q$. (b) The architecture of the Fourier layer in (a). The output $v_0(x)$ from network $P$ is Fourier-transformed into the spectral space, where a spectral weight $R$ is applied along with rectangular truncation. After an inverse Fourier transform, the result is added to $v_0(x)$ and passed through an activation function.}
\label{vanilla_fno}
\end{figure*}

Prior to introducing our proposed network architecture, we first provide a concise overview of the vanilla Fourier Neural Operator (the original FNO without modifications) \cite{FNO_Origin_2021}. The vanilla FNO architecture is designed to learn mappings between infinite-dimensional function spaces for solving PDEs. It is inspired by kernel integral operators for solving PDEs, analogous to Green's function-based convolutions. For spatial problems, by leveraging Fourier transforms, the vanilla FNO translates spatial convolutions into spectral multiplications, effectively parameterizing the integral kernel through learnable linear transformations in the spectral domain. Fig.~\ref{vanilla_fno}(a) illustrates the full architecture of the vanilla FNO, while Fig.~\ref{vanilla_fno}(b) shows the structure of the Fourier layer used in (a). In detail, the input function $a\left(x\right)\in R^{d_{in}}$ is first lifted to a high-dimensional latent space ${v_0\left(x\right)\in R}^w$ via a shallow fully connected network $P$. Each subsequent Fourier layer processes its corresponding input function through: (1) a Fourier transform projecting the input function to spectral space, (2) a linear transformation via learnable spectral weight $R_m$, applied within a truncated spectral band retaining only low-frequency modes, (3) an inverse Fourier transform reconstructing spatial function, (4) a residual addition combining the layer input function and the output function from step (3), and (5) a nonlinear activation function (ReLU) applied to the residual-summed function. The output function of the Fourier layer is given by
\begin{equation}
  \small
v_{m+1}\left(x\right)=\sigma\left(\mathcal{F}^{-1}\left(R_m\odot\mathcal{F}\left(v_m\left(x\right)\right)\right)+W_mv_m\left(x\right)\right),
\end{equation}
where m denotes the index of the Fourier layer and $W_m$ is a learnable linear transformation for the $m$-th layer. The symbol $\odot$ denotes the Hadamard product, representing element-wise multiplication. After Fourier layers, $v_n\left(x\right)$ is projected to the target function $u(x) \in R^{d_{out}}$. Benefiting from spectral-domain transformations, the vanilla FNO can effectively capture global spatial dependencies, making it particularly advantageous for solving PDEs compared to conventional convolution-based methods.

Our model utilizes a Data-Aware Fourier Neural Operator (DA-FNO) within an auto-regressive framework to iteratively predict time-evolving electromagnetic fields. The main difference between the DA-FNO and the vanilla FNO is the design of the Fourier layer, which will be discussed in detail later.

\begin{figure*}[t!]
\centering
\includegraphics[width=15cm]{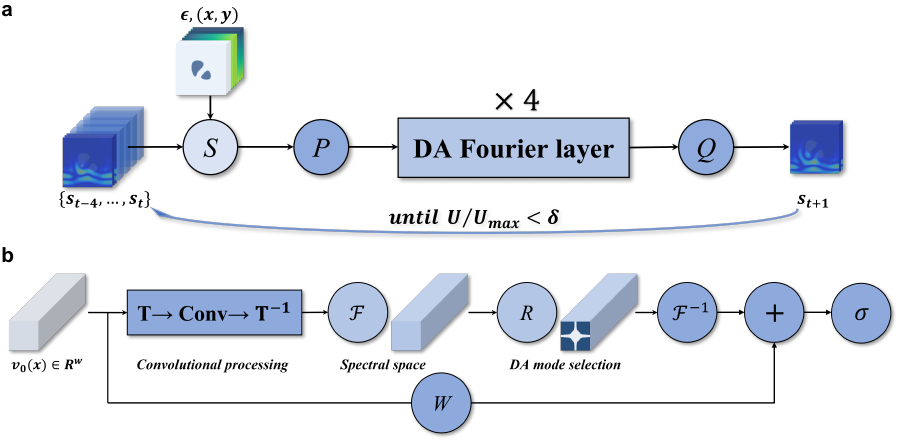}
\caption{(a) The full architecture of the DA-FNO model. The five input time-domain field states $\left \{ s_{t-4},\ldots,s_t \right \}$ is augmented by $S$ with permittivity distribution $\epsilon$ and spatial coordinates $(x,y)$, lifted to a higher-dimensional space by network $P$, processed through four DA Fourier layers and mapped to the next state $s_{t+1}$ by network $Q$. The new state and the last four time-domain states of the input form a new input $\left \{ s_{t-3},\ldots,s_{t+1} \right \}$, enabling the iteration to continue until the energy of the new state falls below a convergence factor, $\delta$, of the historical maximum. (b) The architecture of the DA Fourier layer. The convolutional processing establishes correlations among the three field components. Data-Aware mode selection is performed in the spectral space.}
\label{DA-FNO}
\end{figure*}

A pictorial representation of our model architecture is presented in Fig.~\ref{DA-FNO}(a). The input tensor of the model stacks five consecutive time-domain states $\left \{ s_{t-4},\ldots,s_t \right \}$ along the channel dimension, with each state comprising electromagnetic field components $\left(E_z,H_x,H_y\right)$ for the TM-polarized wave. Spatial coordinates $\left(x,y\right)$ and the permittivity distribution $\epsilon$ are first appended to the channels to encode geometric priors. Next, the augmented input is lifted into a high-dimensional latent space via a fully connected layer $P$, and subsequently transformed through four Data-Aware Fourier layers. Then, a linear layer $Q$ maps the refined latent features to the output time-domain state $s_{t+1}$, representing the electromagnetic field updates. Finally, the oldest state in the input tensor $s_{t-4}$ is discarded, and the predicted state $s_{t+1}$ is appended to form the updated input tensor $\left \{ s_{t-3},\ldots,s_{t+1} \right \}$ for the next iteration. This auto-regressive process continues, with the electromagnetic energy of each newly predicted state being computed as
\begin{equation}
U_{t+1}=\int_{S}{\frac{1}{2}\left(\mathbf{E}\cdot\mathbf{D}+\mathbf{H}\cdot\mathbf{B}\right)dS},
\end{equation}
where $S$ is the simulation domain. The simulation autonomously terminates when $U_{t+1}<\delta\cdot U_{max}$, and $U_{max}$ denotes the maximum energy recorded throughout the simulation history.

\begin{figure*}[t]
\centering
	\includegraphics[width=16cm]{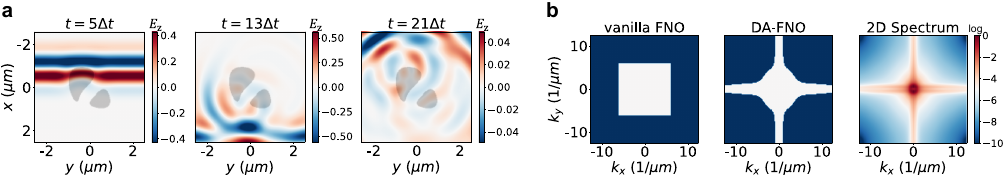}
\caption{(a) Ez field distributions at $t=5\Delta t,13\Delta t $  and  $21\Delta t$. (b) The mode selection regions (white areas) of the vanilla FNO and DA-FNO. Average spectral image of all training data, normalized and displayed in logarithmic scale.}
\label{DA mode selection}
\end{figure*}

Compared to the vanilla FNO model, the DA-FNO model processes input data with an additional dimensionality, which is used to represent the three field components $\left(E_z,H_x,H_y\right)$. These components are inherently coupled through the underlying physical laws according to Maxwell's curl equations without sources:
\begin{equation}
\centering
\begin{aligned}
\nabla\times \mathbf{H} &= \frac{\partial \mathbf{D}}{\partial t} \\
\nabla\times \mathbf{E} &= -\frac{\partial \mathbf{B}}{\partial t}.
\end{aligned}
\end{equation}

However, when directly fed into the vanilla FNO, these physically coupled field components are treated as decoupled channels during spectral processing. As a result, the vanilla FNO fails to capture the intrinsic physical correlations and constraints dictated by the curl equations, which link the evolution of one component to the spatial variations of the others. To address this architectural limitation and enforce physics-compliant interactions, we introduce a learnable $3\times3$ convolution layer with 3 input channels and 3 output channels before the Fourier transform $\mathcal{F}$ in each Fourier layer, as shown in Fig.~\ref{DA-FNO}(b). Here, $T$ denotes a tensor shape transformation that maps the dimension of the field component to the dimension of the feature, while $T^{-1}$ represents its inverse. The convolution takes all three field components as input channels and generates three output components correspondingly. This design allows each field component to dynamically incorporate information from the other two through localized spatial mixing. We adopt convolution rather than other linear transformations because it effectively captures local spatial dependencies, closely resembling the computations of curl operators. Combined with the Fourier transform, the DA-FNO enables complementary learning of both local and global spatial interactions, where local interactions reflect fine-scale updates, and global interactions describe the overall long-term field evolution, according to the FDTD update equations\cite{taflove_fdtd_chapter_2005} (see Supplementary material, Section~1 for more details).

Furthermore, we modify the spectral mode selection strategy to better capture the underlying physics of the scattering process. The vanilla FNO architecture adopts a rectangular truncation window, $\left[-k_{max},k_{max}\right]^2$, to retain low-frequency spectral modes. While effective for suppressing noise, this fixed, low-pass filtering is suboptimal for complex electromagnetic simulations. As a wave interacts with sub-wavelength features in a complex geometry, physical phenomena such as light scattering and diffraction generate fine spatial features in the field distribution. In the spectral domain, accurately representing these features requires retaining the corresponding high-frequency components (large k-vectors). As illustrated in Fig.~\ref{DA mode selection}(a), the $E_z$ field distributions at $t=5\Delta t$, $13\Delta t $, and $21\Delta t$ exhibit progressively intricate spatial patterns, confirming the emergence of these critical high-frequency components over time. To address this, we propose a physically-motivated, data-aware spectral mode selection scheme. Instead of a fixed rectangular window, our method learns the most significant spectral modes directly from the training data by performing three steps: (1) Perform 2D Fourier transforms of spatiotemporal field slices $(E_z,H_x,H_y)$ across the training dataset. (2) Apply min-max normalization to each spectrum to mitigate amplitude decay effects arising from time-domain evolution. (3) Average the normalized spectra and select spectral modes in descending order of magnitude until the cumulative sum exceeds a predefined threshold $\theta$ of the total integrated magnitude. The indices of the selected modes are recorded and used to define the sampling region in each Fourier layer. As shown in Fig.~\ref{DA mode selection}(b), this selection region (white area) for the DA-FNO ($\theta=0.9$) deviates significantly from the rectangular window, showing a much better match to the true spectral distribution of the fields. This allows it to capture more high-amplitude spectral modes crucial for accurately resolving the scattering effects.

\begin{table*}[t]
  \caption{ARL1E of the Six Models ($mean \pm std$)}
  \label{tab:comparison}
  \centering
  \renewcommand{\arraystretch}{1.2}  
  \setlength{\tabcolsep}{12pt}
  \begin{tabular}{l c c c}
    \hline
    Model & Number of selected modes & Train   & Test \\
    \hline
    DA-FNO ($\theta=0.9$) & 1817  & 0.108 $\pm$ 0.017 & 0.235 $\pm$ 0.027 \\
    DA-FNO ($\theta=0.8$) & 933   & 0.124 $\pm$ 0.015 & 0.244 $\pm$ 0.030 \\
    DA-FNO ($\theta=0.7$) & 481   & 0.128 $\pm$ 0.006 & 0.249 $\pm$ 0.018 \\
    DA-FNO ($\theta=0.6$) & 231   & 0.134 $\pm$ 0.007 & 0.253 $\pm$ 0.030 \\
    Conv                  & 1922  & 0.116 $\pm$ 0.011 & 0.246 $\pm$ 0.040 \\
    vanilla               & 1922  & 0.221 $\pm$ 0.021 & 0.970 $\pm$ 0.057 \\
    \hline
  \end{tabular}
\end{table*}

Both the introduction of convolutional processing and this refinement of the mode selection strategy are motivated by the underlying physics of electromagnetic wave propagation. The convolutional layer explicitly models the local coupling between field components as dictated by Maxwell's curl equations. The data-aware mode selection, in turn, ensures that the model preserves the high-frequency spectral information necessary to represent the fine spatial details generated by scattering and diffraction. These physically-motivated design choices collectively define our improved FNO, which we refer to as the Data-Aware Fourier Neural Operator (DA-FNO), as its architecture is explicitly informed by the two physical properties embedded within the simulation data. In addition, we use the Scaled Exponential Linear Unit (SELU)\cite{selu} activation function instead of ReLU in the Fourier layer, as our experiments show it helps reduce prediction errors. A more comprehensive description of the network architecture is provided in the Supplementary material, Section~2.

\begin{figure*}[t]
\centering
	\includegraphics[width=500pt]{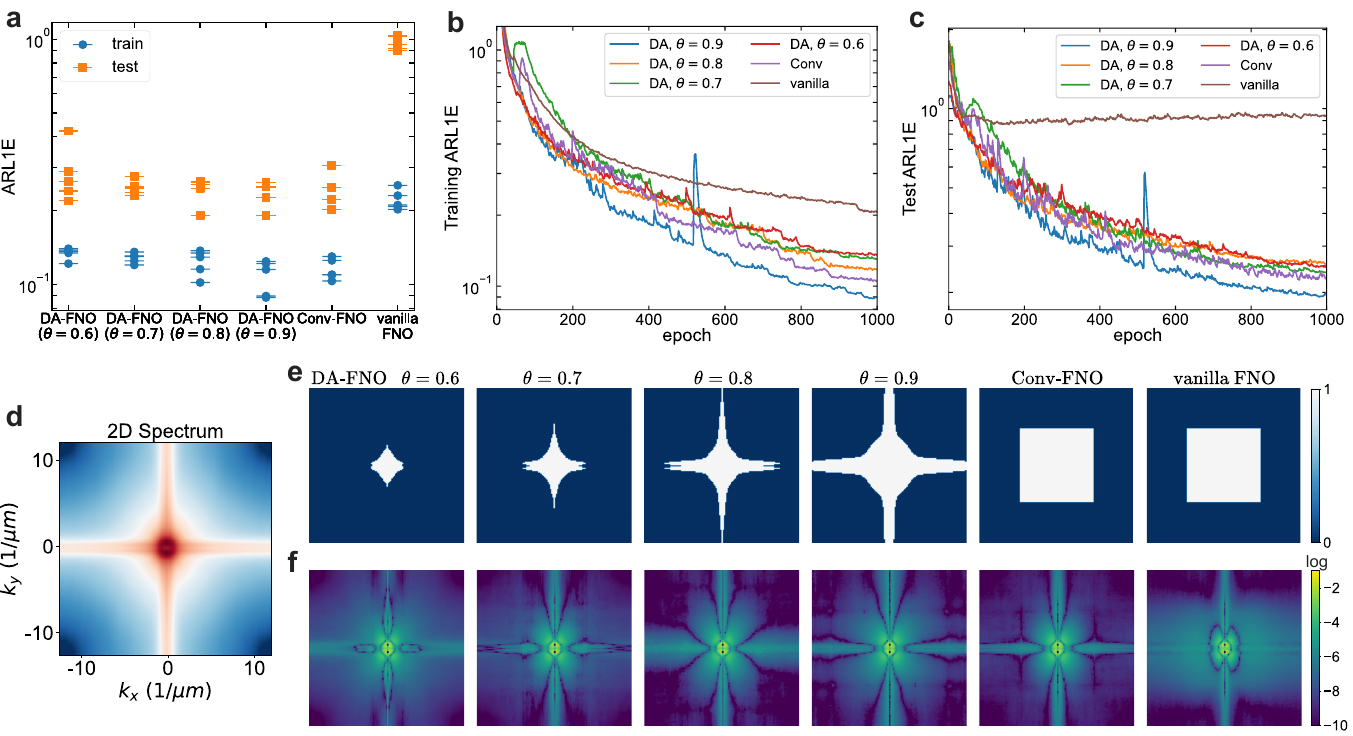}
\caption{Performance comparison among DA-FNO models with $\theta=0.6,0.7,0.8,0.9$, Conv-FNO and vanilla FNO models, based on 500 samples and 1000 epochs. (a) Training and test ARL1Es at the 1000th epoch of the six models. Each model is evaluated over multiple runs, and the five most stable results are presented. (b) Training and (c) test ARL1E curves (smoothed for clarity) of the six models, each corresponding to the run with the lowest training ARL1E among the five runs in (a). All original curves are provided in Supplementary material, Figure S2 and S3. (d) Average spectral image over the training subset. (e) Mode selection regions of the six models. (f) Absolute errors between the average spectral images of the outputs from the best run of each model in (a) and the spectral image of the training subset.}
\label{comparison}
\end{figure*}

\section{Results and Discussion}
The primary objective of the DA-FNO model is to predict the time-domain evolution of electromagnetic fields, supporting photonic simulations across the optical C-band. We generate datasets using an in-house FDTD solver with a time-step of $\Delta t$ under a Gaussian pulse excitation centered at $1550\text{nm}$, with randomized geometries for each sample. The DA-FNO model is trained under three different coarse time-steps $m\Delta t = 12\Delta t, 15\Delta t, \text{and }20\Delta t$, while maintaining a convergence factor of $\delta = 10^{-4}$ for all cases. Further details are provided in Appendix A and Appendix B.
To evaluate the performance on training and testing datasets, we use the Average Relative $L_1$ Error (ARL1E), defined as
\begin{align}
e_{c,t}^{(n)} &= 
\frac{\sum_{i,j} \big| \hat{y}_{c,t}^{(n)}(i,j) - y_{c,t}^{(n)}(i,j) \big|}
     {\sum_{i,j} \big| y_{c,t}^{(n)}(i,j) \big|}, \\[6pt]
\mathrm{ARL1E} &= 
\frac{1}{3NT} \sum_{n=1}^{N} \sum_{t=1}^{T} \sum_{c=1}^{3} e_{c,t}^{(n)},
\end{align}
where $n$,$t$ and $c$ denotes the sample index, time index and component index, respectively. The ARL1E will be used throughout the following analysis. During testing, the model operates in training mode to maintain consistent batch normalization behavior, while no gradient updates are applied. This approach is commonly used when training with small batch sizes due to limited GPU memory (see Supplementary material, Section~4).

\subsection{Performance comparison}
We compare the performance of six models---the DA-FNO models with threshold values $\theta=0.9,0.8,0.7,0.6$, Conv-FNO model and vanilla FNO model. The Conv-FNO model refers to the vanilla FNO with the convolutional processing module from DA-FNO, while retaining the original rectangular mode selection strategy. The number of selected modes in each model is presented in Table~\ref{tab:comparison}.

Through several training runs, we observe that initial conditions have a significant impact on the training dynamics and final performance of the models. Given this sensitivity, we conduct repeated training trials with random weight initializations to ensure that the evaluation results are representative and not biased by outlier cases. For each model, these trials are performed using the first 500 samples of the full dataset, which are partitioned into a training subset (samples 1-450) and a test subset (samples 451-500). Among all trials, five runs with stable and consistent loss descent curves are selected for comparison. From these, the best-performing run of each model is further identified to compare their loss curves and average spectral images. All models are trained for 1000 epochs under identical hyperparameter settings.

The results are presented in Fig.~\ref{comparison} and Table~\ref{tab:comparison}. Fig.~\ref{comparison}(a) shows the training and test ARL1Es at the 1000th epoch of the six models, while Table~\ref{tab:comparison} reports their mean and standard deviation over the five runs. For DA-FNO models, increasing the threshold $\theta$ leads to a consistent downward trend in both training and test errors, indicating that retaining more spectral modes improves the model's ability to simulate electromagnetic fields. At the same time, the training standard deviation increases with $\theta$, as retaining more spectral modes introduces higher-frequency components and increases the model's flexibility. When $\theta$ exceeds 0.9, the remaining modes (as shown in the spectral distribution of Fig.~\ref{DA mode selection}(b)) exhibit very low amplitudes, and even a slight increase in $\theta$ leads to a substantial rise in the number of selected modes. Therefore, we do not consider higher values of $\theta$. Compared with the Conv-FNO model, the DA-FNO ($\theta=0.9$) model achieves slightly lower minimum errors on both training and test sets in the Fig.~\ref{comparison}(a--c), suggesting that the data-aware mode selection strategy is more effective than the rectangular strategy in capturing long-term electromagnetic dynamics. Additionally, both DA-FNO and Conv-FNO models significantly outperform the vanilla FNO model, especially in terms of test error, demonstrating that the incorporation of convolutional processing greatly enhances both model accuracy and generalization.

Fig.~\ref{comparison}(d) shows the average spectral image over the training subset, where higher intensities are primarily concentrated in a central, cross-shaped region. The horizontal and vertical axes are labeled as $k_x$ and $k_y$, respectively, representing the spatial frequencies in the two directions. In Fig.~\ref{comparison}(f), we calculate the absolute errors between the average spectral images of the outputs from the best run of each model and the spectral image of the training subset (Fig.~\ref{comparison}(d)). The errors are found to be closely associated with the mode selection regions depicted in Fig~\ref{comparison}(e). For instance, the mode selection region of the DA-FNO model with $\theta=0.8$ excludes the high-frequency components along the $k_x$ axis, whereas that of $\theta=0.9$ includes them. Consequently, the spectral errors of the $\theta=0.8$ model are larger in the high-frequency region of $k_x$ compared with $\theta=0.9$. Similarly, at the high-frequency ends of $k_y$ along the line $k_x=0$, the errors are highly sensitive to whether this region is covered by the mode selection, a consistent trend observed across all six models.

\subsection{Training results}

\begin{figure*}[!htbp]
\centering
	\includegraphics[width=16.5cm]{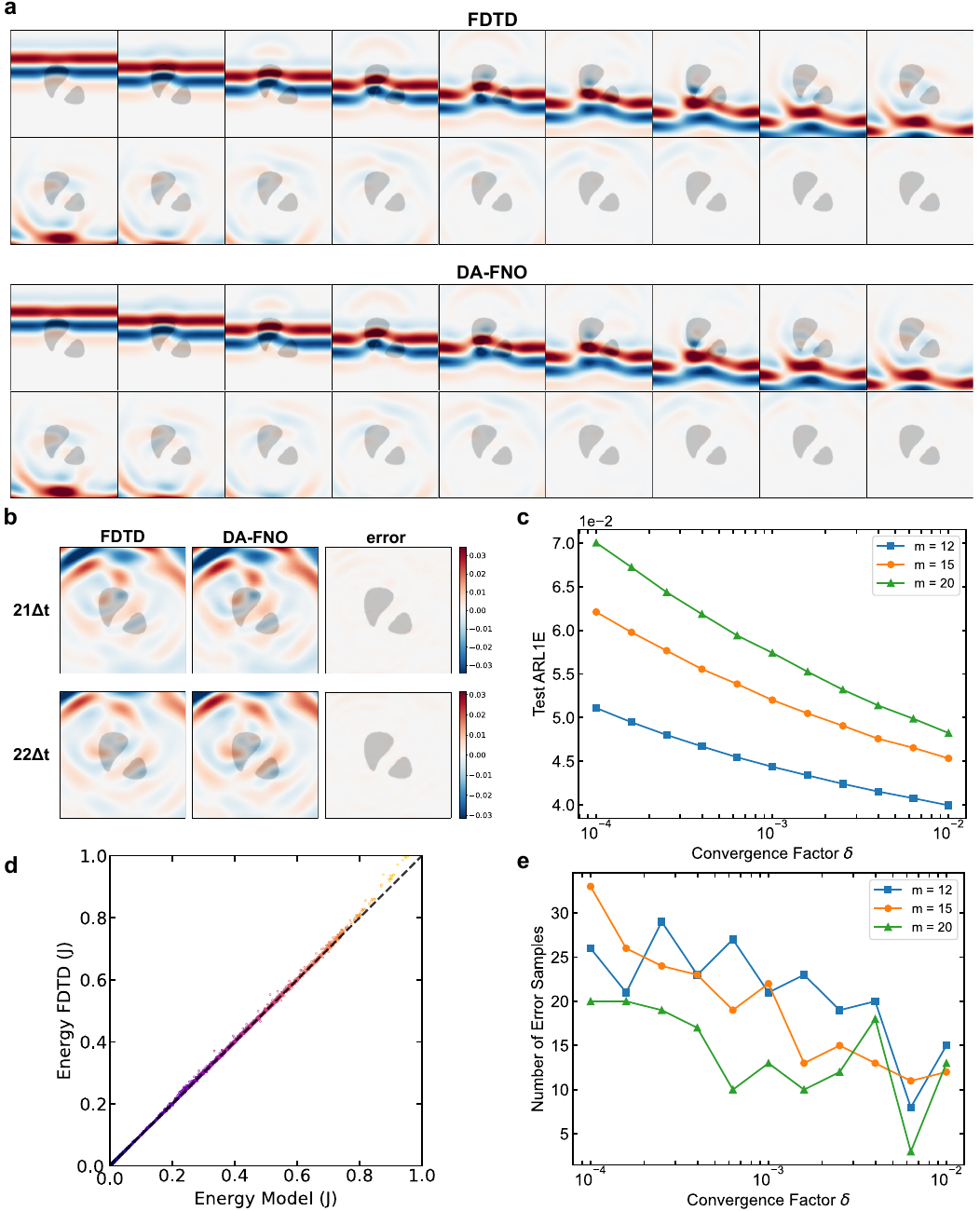}
\caption{(a) Comparison between partial outputs ($H_y$, from $t=5\Delta t$ to $22\Delta t$) of the DA-FNO model and FDTD for a single sample. (b) The last two time-step results in (a) with individually tailored colormaps. (c) Test ARL1E of the DA-FNO model under different coarse time-step $m$ and convergence factor $\delta$. (d) Comparative scatter plot of the normalized energy for all test samples of the DA-FNO model ($m=15$) and FDTD. The black dashed line represents $y=x$. (e) Number of erroneous samples caused by mismatched termination time steps out of 300 tested samples. All errors differ from the ground truth by only one single time-step.}
\label{field plot}
\end{figure*}

Based on the above experiment, the DA-FNO model with $\theta = 0.9$ is employed and trained on the entire training dataset under three different coarse time-steps, $m\Delta t = 12\Delta t, 15\Delta t, \text{and }20\Delta t$, with the convergence factor fixed at $\delta = 10^{-4}$. The final training ARL1Es under the three coarse time-steps are 3.86\%, 4.35\% and 4.53\% respectively. A partial output (from $t=5\Delta t$ to $t=22\Delta t$) of one sample with a relative $L_1$ error of 4.95\% under $m = 15$ is illustrated in Fig.~\ref{field plot}(a). A TM-polarized plane wave interacts with the scatterer upon incidence. As the wave propagates out of the simulation domain, the energy within the region gradually decays. It can be observed from the figure that the results obtained by FDTD and the DA-FNO model are highly similar. For clearer visual comparison, the last two time-step results shown in Fig.~\ref{field plot}(a) are re-visualized in Fig.~\ref{field plot}(b) using individually tailored colormaps. Despite the presence of numerous high-frequency variations in the figure, the outputs of FDTD and the model exhibit strong similarity. Furthermore, we evaluated the model's performance on the test set under different convergence factors, $\delta$, ranging from $10^{-4}$ to $10^{-2}$, as shown in Fig.~\ref{field plot}(c). As $\delta$ increases, the ARL1E decreases. This behavior arises because the DA-FNO model performs iterative field predictions, where errors accumulate at each step; thus, terminating the simulation earlier with a larger  $\delta$ results in smaller accumulated errors. In addition, smaller coarse time-steps result in lower ARL1E, because the field varies less between adjacent steps, making the predictions easier for the model. 

\begin{figure*}[!htbp]
\centering
	\includegraphics[width=15cm]{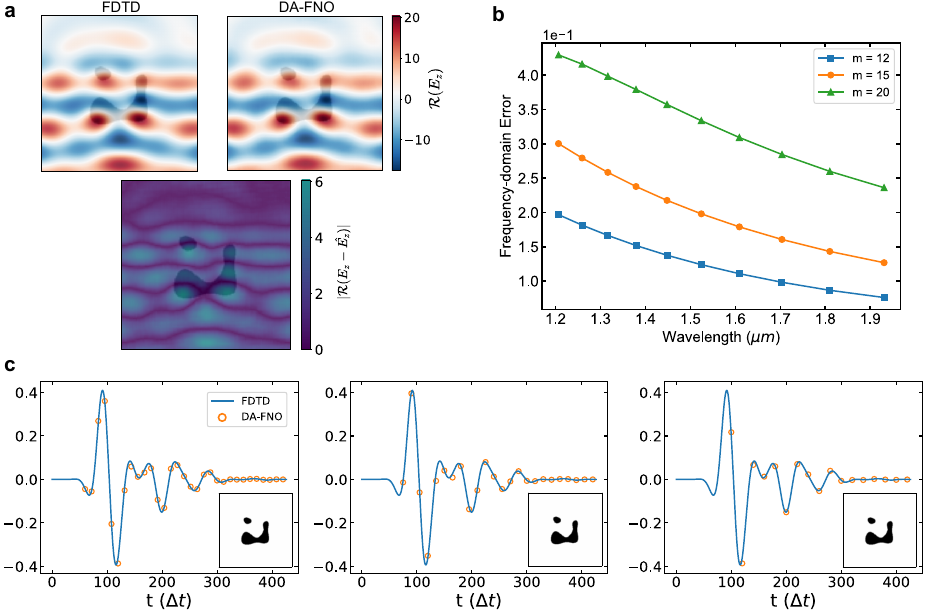}
\caption{(a) Two-dimensional frequency-domain profiles of the $\Re(E_z)$ component for a typical sample, including the FDTD result, the DA-FNO prediction, and their difference. (b) Frequency-domain errors between the FDTD and DA-FNO, evaluated over the wavelength range of $1.2 \mu m$--$1.9\mu m$. (c) Time-domain $E_z$ waveforms at the central point of the simulation domain predicted by the FDTD and the DA-FNO model with $m=12,15,\text{and }20$ from left to right. The subplot in the bottom-right corner is the scatterer geometry.}
\label{frequency_information}
\end{figure*}

The DA-FNO model iteratively outputs fields until energy convergence is detected, at which point the simulation is automatically terminated. To ensure that the termination time determined by the model is correct, we compare the normalized energy at each time step of the outputs from the model and FDTD across all 300 test samples. Fig.~\ref{field plot}(d) presents the results for the case of $m = 15$, while the cases of $m = 12$ and $m = 20$ show similar behavior (see Supplementary material, Figure S4). Each point represents a specific time step of a test sample. Most points lie closely along the reference line $y=x$, indicating good agreement between predicted and true energies. Only a small number of points fall within the high-energy regime, and these points deviate from the reference line. Investigation shows these outliers originate from samples with very small scatterers. When the plane wave interacts with these scatterers, only a very small amount of energy leaks out of the simulation domain, unlike typical cases involving larger dissipation. Consequently, the model follows the typical pattern observed in most cases and underestimates the energy in the simulation domain for these small-scatterer samples. In addition, we compare the termination time steps of the DA-FNO model and FDTD across the 300 test samples. For the vast majority of samples, the two are identical, while fewer than ten percent exhibit a discrepancy of only one time-step. Fig.~\ref{field plot}(e) shows the number of erroneous samples under different values of $m$ and $\delta$. As $\delta$ increases, the number of erroneous samples shows a decreasing trend, because the model's prediction error becomes smaller with a larger $\delta$. And a larger $m$ leads to greater energy differences between adjacent time steps, which reduces the probability of incorrect termination time steps. When calculating the ARL1E, if the termination time steps of model and FDTD are different, we use the smaller one for the calculation. Considering that fewer than ten percent of the samples exhibit a one-time-step discrepancy, this approach has a negligible impact on the accuracy of the error evaluation.

\subsection{Frequency-domain information}
The DA-FNO model outputs the time-domain evolution of the electromagnetic field. By applying a Fourier transform to these time-domain information, we obtain the corresponding frequency-domain field profiles. Fig.~\ref{frequency_information}(a) presents the two-dimensional frequency-domain profile of the $E_z$ component for a typical sample. To evaluate the accuracy of the frequency-domain information, we calculate the frequency-domain field profiles predicted by the DA-FNO model for all test samples and compare them with the corresponding FDTD results. The error is calculated using:
\begin{equation}
  error = \frac{1}{N}\sum_{N}^{} \frac{ {\textstyle \sum_{c,i,j}^{}}\left | \hat{y}(c,i,j) - y(c,i,j) \right |  }{ {\textstyle \sum_{c,i,j}^{}} \left | y(c,i,j) \right | },
\end{equation}
where $c,i,j$ denotes the field component index and x-, y-axis indices. Fig.~\ref{frequency_information}(b) shows the frequency-domain errors of three DA-FNO models, evaluated over the spectral range corresponding to wavelengths from $1.2\mu m$ to $1.9\mu m$, which includes the excitation's central wavelength at $1550\text{nm}$ and captures the region of highest spectral energy. A smaller $m$ and a larger wavelength both lead to lower errors. The output of the DA-FNO model can be interpreted as a sampled version of the FDTD results, as illustrated in Fig.~\ref{frequency_information}(c), which shows the time-domain $E_z$ waveform at the central point of the simulation domain for $m=12,15,\text{and }20$ from left to right. A smaller $m$ corresponds to a finer sampling interval, which yields a more accurate frequency spectrum. And longer-wavelength (lower-frequency) components are less susceptible to spectral aliasing, resulting in lower errors. From the perspective of the time-domain waveform, both a smaller $m$ and the presence of longer-wavelength components allow the sampled points to more closely match the original waveform. This experiment is conducted under the convergence factor $\delta$ of $10^{-4}$. However, varying $\delta$ from $10^{-4}$ to $10^{-2}$ has a negligible impact on the experimental results, because the energy decay between these two levels is minimal, resulting in virtually no difference in the frequency domain.


\subsection{Wavelength generalization}
\begin{figure}[t]
\centering
	\includegraphics[width=.8\columnwidth]{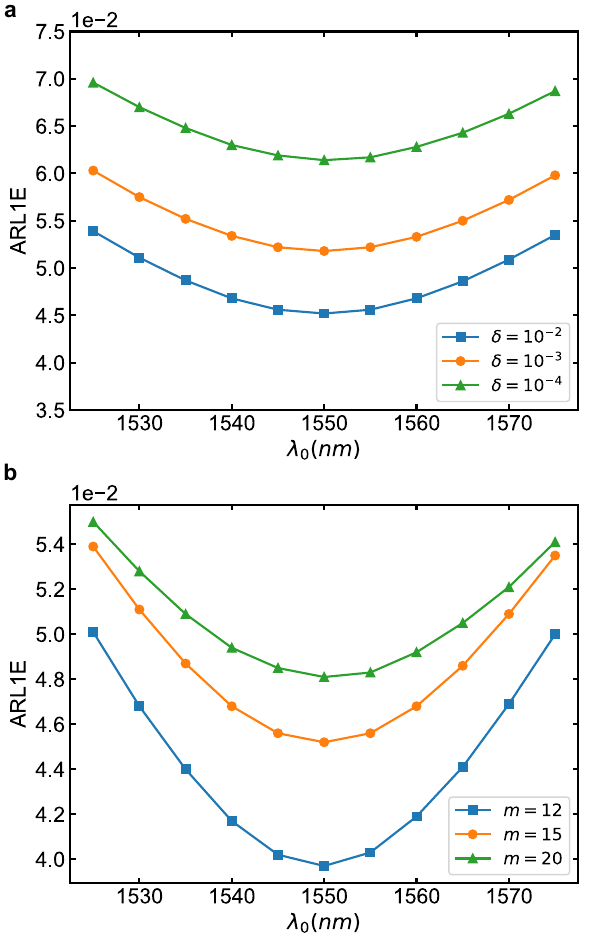}
\caption{Wavelength generalization performance over the $1525\text{nm}$--$1575\text{nm}$ range: (a) results for different values of convergence factor $\delta$ at coarse time-step $m = 15$; (b) results for different values of coarse time-step $m$ at convergence factor $\delta = 10^{-2}$.}
\label{wavelength_generalization}
\end{figure}

We evaluate the wavelength generalization capability of the DA-FNO model on a series of test sets with the central wavelength $\lambda_0$ of the excitation ranging from $1525 \text{nm}$ to $1575 \text{nm}$ with a $5 \text{nm}$ interval. This range covers the entire optical C-band. And we investigate the influence of the convergence factor $\delta$ and the coarse time-step $m\Delta t$ on this generalization performance. Fig.~\ref{wavelength_generalization}(a) shows the model's wavelength generalization performance for different values of $\delta$ at $m = 15$, while Fig.~\ref{wavelength_generalization}(b) presents the performance for different values of $m$ at $\delta=10^{-2}$. In both cases, the ARL1Es increase as the central wavelength moves away from the training wavelength of $1550 \text{nm}$. For Fig.~\ref{wavelength_generalization}(a), the wavelength generalization performance is similar across different values of $\delta$. Throughout the entire C-band, the relative increase in ARL1E with respect to the training wavelength of $1550 \text{nm}$ remains below 11.3\%. In Fig.~\ref{wavelength_generalization}(b), however, different values of $m$ exhibit noticeable effects on generalization. Smaller $m$ leads to a larger increase in ARL1E, indicating poorer wavelength generalization. Section 3.3 has shown that smaller $m$ yields smaller frequency-domain errors. This implies that models with smaller $m$ have higher spectral resolution and are therefore more sensitive to wavelength variations. Consequently, changes in the central wavelength have a greater impact on models with smaller $m$. Although the generalization performance varies with $m$, the relative increase in ARL1E does not exceed 12\% in all cases, and the prediction accuracy across the entire C-band stays around 95\%.

\subsection{Preliminary 3D Investigation}
\begin{figure}[t]
\centering
	\includegraphics[width=.8\columnwidth]{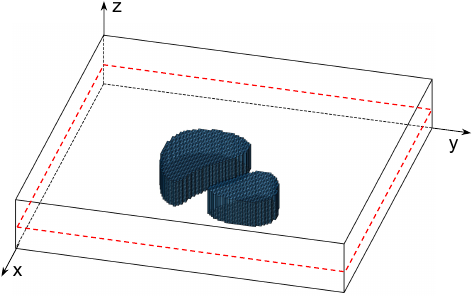}
\caption{Schematic diagram of the three-dimensional extruded simulation configuration. The scattering structure represents a 2D randomized geometry extruded along the $z$-axis. The red dashed rectangle indicates the central cross-sectional plane where the electromagnetic field components ($E_z,H_x,H_y$) are extracted from the 3D FDTD simulation data to evaluate the applicability of the DA-FNO model.}
\label{3D Investigation}
\end{figure}

To further evaluate the applicability of the proposed DA-FNO framework beyond purely two-dimensional settings, we consider a three-dimensional electromagnetic case study based on an extruded geometry. As shown in Fig.~\ref{3D Investigation}, the scatterer exhibits a finite thickness along the $z$ direction, while remaining structurally invariant along this extrusion axis. Owing to this uniformity along the extrusion direction, field distributions on a representative cross-sectional plane provide a meaningful basis for comparison with the two-dimensional DA-FNO predictions. Therefore, we extract the field components ($E_z, H_x, H_y$) on the central plane along the $z$ direction from the generated three-dimensional simulation data and use them for model training and testing. In the three-dimensional setting, electromagnetic energy may resonate along the $z$ direction, making it unreliable to infer the total energy evolution of the entire domain from a single cross-sectional plane. Consequently, instead of using the energy convergence termination criterion, we fix the prediction horizon to 15 time slices with the coarse time-step $m\Delta t = 15 \Delta t$.

The DA-FNO model achieves a low training error of 3.6\% and a test error of 3.95\%. Notably, these values are even lower than the purely two-dimensional benchmarks. This improvement is primarily attributed to the fixed prediction horizon of 15 time steps, which is shorter than the typical energy convergence duration, thereby effectively limiting temporal error accumulation. While these results validate the model's high accuracy on the representative cross-section, obtaining complete volumetric data necessitates a full 3D extension of the DA-FNO model. Theoretically, this extension follows consistent architectural principles, essentially requiring the upgrade of Fourier operators and convolutional layers from 2D to 3D. However, given the significantly higher computational cost and GPU memory requirements associated with 3D tensor operations, we identify this full implementation as a primary direction for our future work.

\subsection{Efficiency evaluation}

The DA-FNO model is capable of overcoming the CFL constraint, enabling faster simulations. To evaluate simulation speed, using different numbers of 2D samples, we test the Meep-based FDTD \cite{oskooi_meep_2010} and the DA-FNO model with three different coarse time-steps. All experiments are conducted on the same server, with detailed configurations provided in Table~\ref{tab:server_config}. For conventional FDTD methods, GPU-based parallel simulation of individual samples is inefficient when the computational domain is small, as data transfer overhead between the CPU and GPU can dominate the overall runtime. Therefore, in these experiments, the Meep-based FDTD solver is executed with CPU-based multithreading, and the optimal thread count is empirically found to be 16. To prevent excessive error accumulation, the Meep-based FDTD employ default double precision with material averaging (subpixel smoothing) disabled. In contrast, the neural network-based model naturally supports batch processing due to its architectural design, which makes it well-suited for execution on GPUs. We set the batch size to 200 in the model simulation, configured with single-precision parameters and material averaging enabled.

As shown in Table~\ref{tab:simulation_time}, when evaluating a single sample, the DA-FNO model exhibits a runtime similar to that of Meep-based FDTD. However, the simulation speed of the DA-FNO model improves as the sample size grows up to the batch size of 200, beyond which the speed plateaus. This indicates that the DA-FNO model exhibits greater computational efficiency in large-sample scenarios. In contrast, the simulation speed of the Meep-based FDTD solver is nearly constant for large sample counts, with the only exception being that a single-sample simulation runs slightly slower because multithreading is not fully utilized. In addition, as $m$ decreases, the simulation speed of the DA-FNO model decreases because more iterations are required. This introduces a trade-off with the findings in the previous sections, where smaller $m$ leads to higher prediction accuracy. Nevertheless, because the DA-FNO model overcomes the CFL constraint, it can trade a small amount of accuracy for substantially faster simulations. For example, with $m = 15$, when the sample size exceeds 200, the DA-FNO model achieves a speedup of approximately $11\times$ compared to the multithreaded Meep-based FDTD.

\begin{table}[t]
    \centering
    \caption{Configuration of the server used for all experiments.}
    \label{tab:server_config}
    \begin{tabular}{ll}
        \hline
        Component & Specification \\
        \hline
        CPU   & Intel Xeon Gold 6346 (16 cores) \\
        RAM   & 512 GB DDR4-3200 \\
        GPU   & NVIDIA GeForce RTX 5090 (32 GB) \\
        CUDA  & 12.8 \\
        cuDNN & 9.7 \\
        \hline
    \end{tabular}
\end{table}

\begin{table}[t]
\centering
\caption{Simulation Time (s): DA-FNO vs. FDTD across Sample Sizes}
\label{tab:simulation_time}
\begin{tabular}{lcccc}
\hline
Solver & 1 & 100 & 200 & 300 \\
\hline
DA ($m=12$) & 0.30 & 3.08 & 5.57 & 8.31 \\
DA ($m=15$) & 0.28 & 2.59 & 4.62 & 6.90 \\
DA ($m=20$) & 0.24 & 2.06 & 3.52 & 5.11 \\
Meep FDTD & 0.29 & 25.6 & 50.56 & 75.77 \\
\hline
\end{tabular}
\end{table}
\subsection{Implications for inverse design}

Building upon the demonstrated capabilities, we envision applications of the DA-FNO model in inverse design, where the initial stages often require evaluating a large number of candidate structures, placing a premium on computational efficiency while tolerating moderate accuracy. Most existing works have primarily focused on optimizing devices such as power splitters\cite{power_splitter_LPR_2020,power_splitter_micronmachines_2023}, wavelength demultiplexers\cite{wavelength-division_OE_2022}, scatterers\cite{FNO_ACS_2023}, lenses\cite{PINN_APL_2023}, and metasur\-faces\cite{metaface_ACS_2025}. The performance of these devices is typically characterized in the frequency domain, focusing on spectral responses and steady-state field distributions. In contrast, relatively few studies have addressed devices whose functionalities inherently rely on time-domain characteristics, such as dispersion-engineered components or ultrafast modulators. The modeling of such devices typically relies on time-domain simulations and is often restricted to FDTD-based approaches. The DA-FNO model, by efficiently learning the time-domain evolution of electromagnetic fields, provides a potential pathway to incorporate time-domain dynamics into inverse design frameworks. This capability could enable the design of novel photonic devices where both frequency-domain performance and time-domain behavior are jointly optimized.

\section{Conclusion}

In this work, we propose a novel Fourier neural operator tailored for photonic simulation, termed the Data-Aware Fourier Neural Operator (DA-FNO). The architecture of DA-FNO is guided by physical principles, incorporating convolutional operations to capture the interactions among different components of the electromagnetic field and an adaptive mode selection to preserve the spectral features of scattering phenomena. By embedding the DA-FNO within an auto-regressive framework, we develop a novel neural-operator-based model capable of simultaneously predicting the time-domain evolution of electromagnetic fields and generalizing across complex and randomized geometries. Moreover, it demonstrates robust generalization across complex geometries and the optical C-band, exhibiting a trade-off between accuracy and computational efficiency through the choice of the coarse time-step and the convergence factor. In a representative configuration, it achieves an $11\times$ speedup over conventional FDTD methods with about 95\% accuracy. These results suggest that the DA-FNO model provides a new pathway for C-band photonic simulations. Its ability to accurately capture time-domain dynamics indicates its potential to aid in the inverse design of novel photonic devices, offering a promising direction for physics-aware neural operators in photonics.





\appendix
\renewcommand{\thefigure}{A.\arabic{figure}} 
\setcounter{figure}{0} 
\section{Data generation}
We generate randomized geometries following the meth\-odology of Ref.~\cite{data_generation_OE_2022}. A $256\times256$ matrix is first initialized with uniformly distributed random values between 0 and 1. A 2D Gaussian filter $\left(\sigma=30\right)$ is then applied to the central $250\times250$ region of the matrix, while the peripheral regions remain zero-padded to confine structural features within the domain. The smoothed output is binarized using a threshold of 0.5: values above the threshold are set to 1 (solid material), and those below to 0 (void). To mitigate abrupt material transitions at interfaces, the binary map is down-sampled to a $128\times128$ matrix via $2\times2$ local averaging. This averaging operation smooths interfacial discontinuities while preserving topological randomness, enabling more physically realistic representation of arbitrary geometries (see Supplementary material, Figure S5). We subsequently map the $128\times128$ matrix to a permittivity distribution using a simple mapping:
\begin{equation}
\epsilon\left(r\right)=\left[n_{void}+\left(n_{solid}-n_{void}\right)\chi\left(r\right)\right]^2,
\label{s3}
\end{equation}
where $\chi\left(r\right)\in\left\{0,1\right\}$ denotes the randomized geometry, $n_{void}=1$ and $n_{solid}=1.4$ represent air and material respectively.

\begin{figure}
\centering
	\includegraphics[width=.9\columnwidth]{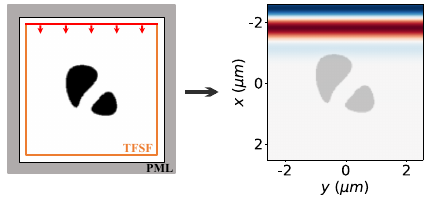}
\caption{Illustration of the data generation process.}
\label{data generation}
\end{figure}

We employ an in-house FDTD solver to perform the simulations and set the convergence factor, $\delta=10^{-4}$. The discrete permittivity distribution $\epsilon\left(r\right)$ is applied within a total-field/scattered-field (TFSF) region, which is enclosed by a perfectly matched layer (PML) to suppress spurious reflections at the domain boundaries. A TM-polarized Gaussian pulsed plane wave $\left(\lambda_0=1550\text{nm},\tau_{FWHM}=30fs\right)$ is launched along the principal axis to simulate the interaction with the randomized scattering geometry, as shown in Fig.~\ref{data generation}. The simulation continues until the domain energy converges $\left(U_t<\delta\cdot U_{max}\right)$. The spatial and temporal resolutions are set to $\Delta x=\Delta y=40\ nm/px$ and $ \Delta t=\frac{1}{c}[(\frac{1}{\Delta x})^2+(\frac{1}{\Delta y})^2]^{-\frac{1}{2}}$ to satisfy the CFL stability condition.

A total of 3,000 simulation samples are generated at a $1550\text{nm}$ central wavelength, of which 2,700 are used for training and the remaining 300 for testing. To evaluate wavelength generalization, we further use the same geometries to generate 10 additional test sets (300 samples each) for central wavelengths ranging from $1525\text{nm}$ to $1575\text{nm}$ at $5\text{nm}$ intervals, excluding the $1550\text{nm}$ training wavelength. This wavelength range covers the optical C-band ($1530\text{-}1565\text{nm}$).

For the three-dimensional case study presented in Section 3.5, we generate 3000 3D samples based on the same randomized geometries described above, comprising 2,700 samples for training and 300 for testing. The 2D binary geometries are extruded along the $z$-axis with a thickness of 8 grids to form the 3D structures, within a simulation domain that extends 15 grids along the $z$-direction. PML boundaries are applied in all three spatial directions to mimic an open environment. A TM-polarized Gaussian plane wave source, identical to the 2D case, is used for excitation. The spatial resolution in the $z$ direction is set to $\Delta z = 40\text{nm}$, consistent with the transverse resolution.

\section{Network training}
The proposed DA-FNO model eliminates the CFL condition, enabling larger time-step iterations while simultaneously generating all components of the electromagnetic field at each iteration. Before training, the model's temporal resolution is set to $m$ times the FDTD time step ($\Delta t_{model}=m\Delta t_{FDTD}$). We sample the full-field data $\left(E_z,H_x,H_y\right)$ from datasets at every $m$th FDTD time step and scale the magnetic field components by the intrinsic impedance of free space, $Z_0=\sqrt{\mu_0 / \epsilon_0}=120\pi\ \Omega$, to match the magnitude of the electric field. Subsequently, each data sample is normalized to the range $\left[-1,1\right]$, motivating the use of a $tanh$ activation function in the model's output layer. These preprocessing steps promote numerical stability, balanced feature scaling, and improved training generalization.

Due to the energy convergence termination criterion in the model simulation, the simulation duration---and consequently the temporal length of the generated data---varies across different geometries. To facilitate batch processing and accelerate training, each sample within a mini-batch is zero-padded to a uniform length $T$, and a mask is defined to record the padded entries for subsequent loss computation. For each mini-batch, the initial five time slices serve as the input $\mathbf{x}$, while the remaining $T-5$ time slices constitute the label data $\mathbf{y}$. The model iteratively predicts the field components for $T-5$ steps, and all predicted components are concatenated to form $\mathbf{\hat{y}}$. The loss function for backpropagation is then computed using $\mathbf{y}$, $\mathbf{\hat{y}}$, and the $\mathbf{mask}$:
\begin{equation}
\mathcal{J}(\mathbf{\hat{y},\mathbf{y},\mathbf{mask}})=\sum_{n,t}(\mathbf{mask} \odot (\frac{1}{3}\sum_{c=1}^{3}\mathcal{L}_2(\mathbf{\hat{y}},\mathbf{y}))),
\end{equation}
where $n$,$t$ and $c$ denotes the sample index, time index and component index, respectively. The $\mathbf{mask}$ is a 2D tensor of 0s and 1s, where 1 indicates valid data and 0 denotes padding, excluding the influence of padded regions during backpropagation. $\mathcal{L}_2(\mathbf{\hat{y}},\mathbf{y})$ represents the relative $L_2$ loss, defined as:
\begin{equation}
{\mathcal{L}_2}_{c,t}^{(n)} =  \frac{\left( \sum_{i,j} \big| \hat{y}_{c,t}^{(n)}(i,j) - y_{c,t}^{(n)}(i,j) \big|^2 \right)^{1/2}}
     {\left( \sum_{i,j} \big| y_{c,t}^{(n)}(i,j) \big|^2 \right)^{1/2}}.
\label{s6}
\end{equation}

The loss function $\mathcal{J}$ computes the relative $L_2$ error across all field components and time steps, thereby preventing the model from discounting later time steps with lower magnitudes and encouraging more consistent prediction accuracy throughout the entire time-domain sequence.

All training was conducted on NVIDIA A6000 and A100 GPUs. More comprehensive description of the training process is provided in Supplementary material, Section~3.

\section*{Data Availability}
The code developed for this study, together with the first 200 samples of the training dataset, is openly available at \href{https://huggingface.co/datasets/Litbeginner/DA-FNO}{\texttt{DA-FNO dataset (HuggingFace)}}
. Additional datasets supporting the findings of this work are available from the corresponding author upon reasonable request.

\section*{Supplementary Material}
Supplementary material associated with this article is provided as a separate file.

\bibliographystyle{elsarticle-num}
\bibliography{sample}

\end{document}